\documentclass[runningheads]{llncs}
\usepackage{graphicx}
\usepackage{comment}
\usepackage{envmath}
\usepackage{amsmath}
\usepackage{multirow}
\usepackage{adjustbox}

\usepackage{xcolor}

\begin{document}
\title{Stability versus Meta-stability in a Skin Microbiome Model}
%
%
\author{Eléa Thibault Greugny\inst{1,}\inst{2} 
\and Georgios N. Stamatas\inst{1} 
\and François Fages\inst{2} 
}
\authorrunning{E. Thibault Greugny et al.}

\institute{Johnson \& Johnson Santé Beauté France, Issy-les-Moulineaux, France \and
Inria Saclay, Lifeware Team, Palaiseau, France
}
\maketitle              
\begin{abstract}
The skin microbiome plays an important role in the maintenance of a healthy skin. It is an ecosystem, composed of several species, competing for resources and interacting with the skin cells. Imbalance in the cutaneous microbiome, also called dysbiosis, has been correlated with several skin conditions, including acne and atopic dermatitis. Generally, dysbiosis is linked to colonization of the skin by a population of opportunistic pathogenic bacteria (for example \textit{C. acnes} in acne or \textit{S. aureus} in atopic dermatitis). Treatments consisting in non-specific elimination of cutaneous microflora have shown conflicting results. It is therefore necessary to understand the factors influencing shifts of the skin microbiome composition. In this work, we introduce a mathematical model based on ordinary differential equations, with 2 types of bacteria populations (skin commensals and opportunistic pathogens) to study the mechanisms driving the dominance of one population over the other. 
By using published experimental data, assumed to correspond to the observation of stable states in our model, we derive constraints that allow us to reduce 
the number of parameters of the model from 13 to 5. 
Interestingly, a meta-stable state settled at around 2 days following the introduction of bacteria in the model, is followed by a reversed stable state after 300 hours.
On the time scale of the experiments,
we show that certain changes of the environment, like the elevation of skin surface pH, create favorable conditions for the emergence and colonization of the skin by the opportunistic pathogen population.
Such predictions help identifying potential therapeutic targets for the treatment of skin conditions involving dysbiosis of the microbiome,
and question the importance of meta-stable states in mathematical models of biological processes.

\end{abstract}

\section{Introduction}
Located at the interface between the organism and the surrounding environment, the skin constitutes the first line of defense against external threats, including irritants and pathogens. In order to control potential colonization of the skin surface by pathogens, the epidermal cells, called keratinocytes, produce antimicrobial peptides (AMPs) \cite{pazgier_human_2006}. The physiologically acidic skin surface pH also contributes to control the growth of bacterial populations \cite{proksch_ph_2018,korting_differences_1990}. Another contributor to the defense against pathogen colonization are commensal bacteria in the community of microorganisms living on the skin, commonly referred to as the skin microbiome. Over the past decade, several studies have highlighted the key role played by such commensal bacterial species defending against invading pathogens, as well as their contribution to the regulation of the immune system \cite{lai_commensal_2009,cogen_staphylococcus_2010,lai_activation_2010,kong_skin_2011,belkaid_dialogue_2014,byrd_human_2018}.

Alterations in the composition of the skin microbiome resulting in a dominance by a pathogenic species, also called dysbiosis, have been associated with skin conditions such as acne or atopic dermatitis (AD) \cite{leyden_propionibacterium_1975,kong_temporal_2012}. In the case of AD, the patient skin is often colonized by \textit{Staphylococcus aureus} (\textit{S. aureus}), especially on the lesions \cite{kong_temporal_2012}. Treatment strategies targeting non-specific elimination of cutaneous microflora, such as bleach baths, have shown conflicting results regarding their capacity to reduce the disease severity \cite{chopra_efficacy_2017}. On the other hand, treatments involving introduction of commensal species, like \textit{Staphylococcus hominis} \cite{nakatsuji_development_2021} on the skin surface appear promising. Accordingly, the interactions between the commensal populations, pathogens and skin cells seem at the heart of maintaining microbiome balance. There is therefore a necessity to investigate further those interactions and the drivers of dominance of one population over others. Unfortunately, it is challenging to perform \textit{in vitro} experiments involving more than one or two different species, even more so on skin explants or skin equivalents. 

Mathematical models of population dynamics have been developed and used for more that 200 years \cite{malthus_essay_1798}. Here, we introduce a model based on ordinary differential equations (ODEs), describing the interactions of a population of commensal species with one of opportunistic pathogens and the skin cells.
We study the factors influencing the dominance of one population over the other on a microbiologically relevant timescale of a couple of days
corresponding to biological experimental data.
More specifically, we identify constraining relationships on the parameter values, based on published experimental data 
\cite{nakatsuji_antimicrobials_2017,kohda_vitro_2021}, corresponding to special cases of our model,
allowing us to reduce 
the parametric dimension of our model from 13 to 5 parameters.
Interestingly, we observe in the reduced model a phenomenon of meta-stability \cite{TK08neuron,RSNGW15cmsb}, also called quasi-stability, 
in which the seemingly stable state reached after 30 hours
following the initiation of the experiment,
is followed after 300 hours by a reversed stable state.
On the time scale of the experiments,
we show that certain changes in the environment, like an elevation of skin surface pH, create favorable conditions for the emergence and colonization of the skin by the opportunistic pathogen population.
Such predictions can help identify potential therapeutic strategies for the treatment of skin conditions involving microbiome dysbiosis,
and underscore the importance of meta-stable states in the real biological processes 
at their different time scales.

\section{Initial ODE model with 13 parameters}
The model built in this paper considers two types of bacterial populations. The first population, $S_c$, regroups commensal bacteria species having an overall beneficial effect for the skin, and the second population, $S_p$, represents opportunistic pathogens.
The differential equations for both bacterial populations are based on the common logistic growth model \cite{zwietering_modeling_1990}, considering non-explicitly the limitations in food and space. The limited resources are included in the parameters $K_{sc}$ and $K_{sp}$, representing the optimum concentration of the populations in a given environment, considering the available resources.

The bactericidal effect of antimicrobial peptides (AMPs) produced by skin cells, $Amp_h$, on $S_p$ is included with a Hill function. This type of highly non-linear functions have been used previously to model the effect of antibiotics on bacterial populations \cite{meredith_bacterial_2015}. For the sake of simplicity, the AMPs produced by skin cells is introduced as a constant parameter, $[Amp_h]$, in the model. It represents the average concentration of these AMPs among surface cells, under given human genetic background and environmental conditions.

Several studies revealed that commensal bacterial populations, like \textit{S. epidermidis} or \textit{S. hominis}, are also able to produce AMPs targeted against opportunistic pathogens, such as \textit{S. aureus} \cite{cogen_selective_2010,nakatsuji_antimicrobials_2017}.
For these reasons, we introduce in the model AMPs of bacterial origin, $Amp_b$, acting similarly to $Amp_h$ on the pathogenic population $S_p$. $Amp_b$ is produced at rate $k_c$ by $S_c$, and degraded at rate $d_a$. Furthermore, we include a defense mechanism of $S_p$ against $S_c$ with a direct killing effect. 

Altogether, this gives us the following ODE system with 3 variables and 13 parameters, all taking non-negative values:

\begin{System}\label{fullSystem}
\frac{d [S_c]}{dt} = \left( r_{sc} \left( 1 - \frac{[S_c]}{K_{sc}} \right) - \frac{d_{sc} [S_p]}{C_1 + [S_p]} \right) [S_c] \\ \\

\frac{d [S_p]}{dt} = \left(r_{sp} \left( 1 - \frac{[S_p]}{K_{sp}} \right) - \frac{d_{spb} [Amp_b]}{C_{ab} + [Amp_b]} - \frac{d_{sph} [Amp_h]}{C_{ah} + [Amp_h]} \right) [S_p]\\ \\

\frac{d [Amp_b]}{dt} = k_c [S_c] - d_a
\end{System}

\begin{table}[t]
\caption{List of the parameters and variables of our mathematical model with their units. CFU = Colony forming unit, AU = Arbitrary Unit, ASU = Arbitrary Surface Unit}\label{tab13param}

\begin{tabular}{|c|p{10cm}|}
\hline
\textbf{Variable} &  \textbf{Interpretation (unit)}\\
\hline
$[S_c]$ & Surface apparent concentration of $S_c$ ($CFU.ASU^{-1}$)\\
$[S_p]$ & Surface apparent concentration of $S_p$ ($CFU.ASU^{-1}$)\\
$[Amp_b]$ & Concentration of $Amp_b$ ($AU.ASU^{-1}$)\\
\hline
\textbf{Parameter} & \textbf{Interpretation (unit)}\\
\hline
$r_{sc}$ & Growth rate of $S_c$ ($h^{-1}$) \\
$r_{sp}$ & Growth rate of $S_p$, ($h^{-1}$) \\
$K_{sc}$ & Optimum concentration of $S_c$ ($CFU.ASU^{-1}$) \\
$K_{sp}$ & Optimum concentration of $S_p$ ($CFU.ASU^{-1}$) \\
$d_{sc}$ & Maximal killing rate of $S_c$ by $S_p$ ($h^{-1}$) \\
$C_1$ & Concentration of $S_p$ inducing half the maximum killing rate $d_{sc}$ ($CFU.ASU^{-1}$) \\
$d_{spb}$ & Maximal killing rate of $S_p$ by $Amp_b$, ($h^{-1}$) \\
$C_{ab}$ & Concentration of $Amp_b$ inducing half the maximum killing rate $d_{spb}$ ($AU.ASU^{-1}$) \\
$d_{sph}$ & Maximal killing rate of $S_p$ by $Amp_h$, ($h^{-1}$) \\
$C_{ah}$ & Concentration of $Amp_h$ inducing half the maximum killing rate $d_{sph}$ ($AU.ASU^{-1}$) \\
$[Amp_h]$ & Concentration of AMPs produced by the skin cells ($AU.ASU^{-1}$)\\
$k_c$ & Production rate of $Amp_b$ by $S_c$ ($AU.h^{-1}.CFU^{-1}$) \\
$d_a$ & Degradation rate of $Amp_b$ ($AU.h^{-1}$)\\
\hline
\end{tabular}

\end{table}

\begin{figure}
    \centering
    \includegraphics[width=\textwidth]{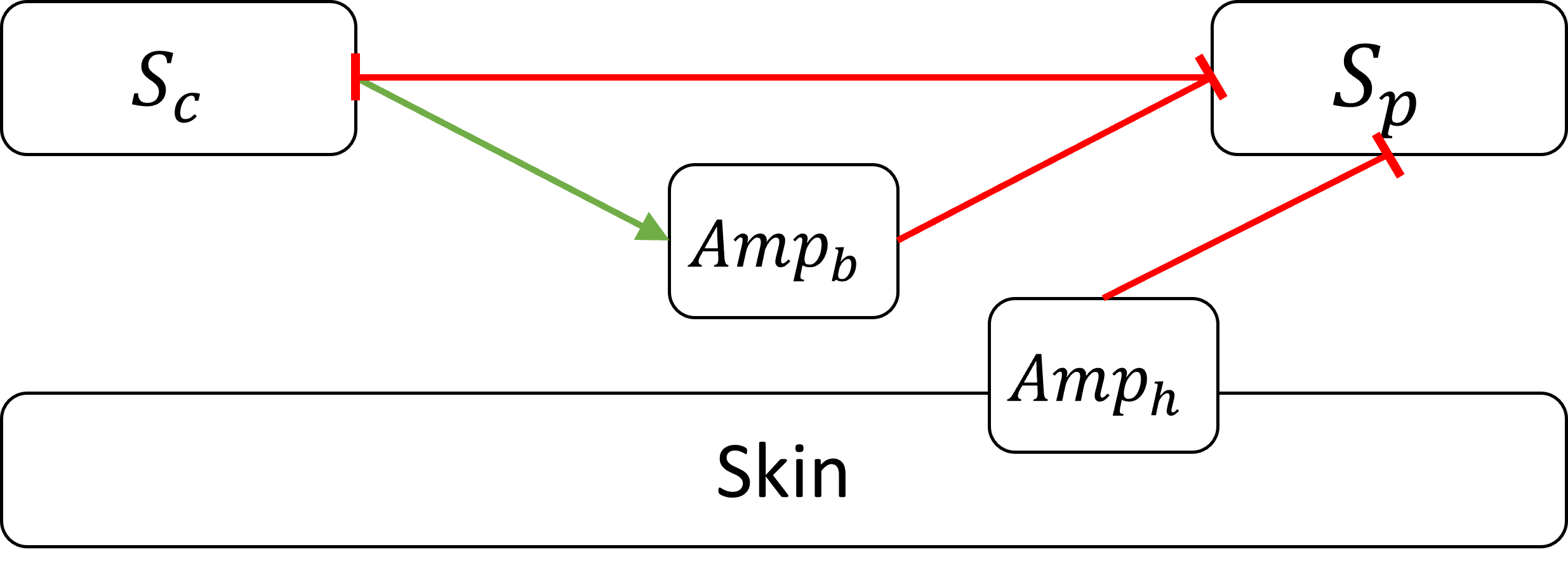}
    \caption{Model overview, green arrow representing production and red T-lines representing killing effect.}
    \label{fig:ModelOverview}
\end{figure}

The model is illustrated on Fig. \ref{fig:ModelOverview} and Table \ref{tab13param} recapitulates the variables and the parameters with their unit.

Such a model cannot be solved analytically. Furthermore, the use of optimization algorithms to infer the 13 parameter values from data resulted in many valid sets of parameter values. Therefore, it is clearly necessary to restrict the number of parameters by identifying some of them, to be able to analyze the model.

\section{Using published experimental data to define relations between model parameters by steady-state reasoning}
The amount of quantitative experimental data available for the model calibration is very limited due to the difficulty of carrying out experiments involving co-cultures of different bacterial species. Most of the published work focuses on single species or on measuring the relative abundances of species living on the skin, which is highly variable between individuals and skin sites \cite{grice_topographical_2009}. In the case of AD specifically, \textit{S. aureus} is considered pathogenic and \textit{S. epidermidis} commensal. Published data exist however for those species which we can use to constrain the parameter values of the model.

Two series of \textit{in vitro} experiments are considered \cite{nakatsuji_antimicrobials_2017,kohda_vitro_2021}. While \textit{in vitro} cultures, even on epidermal equivalent, do not entirely capture the native growth of bacteria on human skin, they provide useful quantitative data that would be very difficult to measure \textit{in vivo}.

In the first experiment \cite{kohda_vitro_2021}, mono-cultures and co-cultures of \textit{S. epidermidis} and \textit{S. aureus} were allowed to develop on a 3D epidermal equivalent. 
Table \ref{tabExpDataKohda} recapitulates the population sizes of the two species measured after 48 hours of incubation. Kohda \textit{et al.} also performed another co-culture experiment where \textit{S.epidermidis} was inoculated 4 hours prior to \textit{S.aureus} in the media. This data is not used here as it requires additional manipulation to match the situation represented by the model. However, it would be interesting to use it in the future for model validation.

In the second experiment \cite{nakatsuji_antimicrobials_2017} the impact of human (LL-37) and bacterial (\textit{Sh}-lantibiotics) AMPs on \textit{S. aureus} survival was studied. The experiments were performed \textit{in vitro}, and the \textit{S. aureus} population size was measured after 24 hours of incubation.
Table \ref{tabExpDataNakatsuji} summarizes their observations.

\begin{table}
\caption{Experimental data from Kohda et. al \cite{kohda_vitro_2021} used for identifying parameter values. 
}\label{tabExpDataKohda}
\begin{center}
\begin{tabular}{l|c|c}
& \textbf{\textit{S. epidermidis} (CFU/well)} & \textbf{\textit{S. aureus} (CFU/well)}\\
\hline
\textbf{Mono-cultures} & $4.10^8$ & $3.10^9$ \\
\textbf{Co-cultures} & $1.10^8$ & $1.10^9$ \\
\end{tabular}
\end{center}
\end{table}

\begin{table}
\caption{Experimental data from Nakatsuji et. al \cite{nakatsuji_antimicrobials_2017} used for identifying parameter relations. 
}\label{tabExpDataNakatsuji}
\begin{center}
\begin{tabular}{c|c|c}
 \textbf{\textit{Sh}-lantibiotics ($\mu M$)} & \textbf{LL-37 ($\mu M$) }& \textbf{\textit{S. aureus} (CFU/mL)}\\
\hline
0 & 4 & $10^9$\\
0 & 8 & $6.10^5$\\
0.32 & 0 & $5.10^8$\\
0.64 & 0 & $3.10^3$
\end{tabular}
\end{center}
\end{table}

\subsection{Parameter values inferred from mono-culture experiment data} \label{subsec:Param_mono}
We consider first the monocultures experiments from Kohda \textit{et al.} \cite{kohda_vitro_2021}, representing the simplest experimental conditions. \textit{S. epidermis} is a representative of the commensal population $S_c$, and \textit{S. aureus} of the pathogenic one, $S_p$. 
Since the two species are not interacting, the set of equations simplifies to:

\begin{System}
\frac{d [S_c]}{dt} = \left( r_{sc} \left( 1 - \frac{[S_c]}{K_{sc}} \right) \right) [S_c] \\ \\

\frac{d [S_p]}{dt} = \left(r_{sp} \left( 1 - \frac{[S_p]}{K_{sp}} \right) \right) [S_p]
\end{System}

At steady-state, the population concentrations are either zero, or equal to their optimum capacities ($K_{sc}$ or $K_{sp}$) when the initial population concentration is non-zero. Given the rapid growth of bacterial population, the experimental measurements done after 48 hours of incubation can be considered as corresponding to a steady-state, which gives:

\begin{equation} \label{Kse}
K_{sc} = 4.10^8 \; CFU.ASU^{-1}
\end{equation}
\begin{equation} \label{Ksa}
K_{sp} = 3.10^9 \; CFU.ASU^{-1}
\end{equation}

\subsection{Parameter relations inferred from experimental data on AMP}
The experimental conditions of Nakatsuji \textit{et al.} \cite{nakatsuji_antimicrobials_2017} correspond to the special case where there is no commensal bacteria alive in the environment, only the bacterial AMPs, in addition to those produced by the skin cells. Our system of equations then reduces to:

\begin{equation}
\frac{d [S_p]}{dt} = \left(r_{sp} \left( 1 - \dfrac{[S_p]}{K_{sp}} \right) - \frac{d_{spb} [Amp_b]}{C_{ab} + [Amp_b]} - \frac{d_{sph} [Amp_h]}{C_{ah} + [Amp_h]} \right) [S_p]
\end{equation}

The concentrations in LL-37 and \textit{Sh}-lantibiotics, translated in our model into $[Amp_h]$ and $[Amp_b]$ respectively, are part of the experimental settings. Therefore, we consider them as constants over time. At steady state, we get:

\begin{equation} \label{NakaEqSS}
[S_p]^* = 0 \quad \textrm{or} \quad [S_p]^* = K_{sp} \left ( 1 - \frac{d_{spb} [Amp_b]}{r_{sp} (C_{ab} + [Amp_b])} - \frac{d_{sph} [Amp_h]}{r_{sp} (C_{ah} + [Amp_h])} \right)
\end{equation}

Let us first focus on the special case where no \textit{Sh}-lantibiotics were introduced in the media, translating into $[Amp_b] = 0$ in our model. We consider again that the biological observations after 24 hours of incubation correspond to steady-state and substitute the experimental values measured \\
$[Amp_h] = 4 \,\mu M$ ; $[S_p]^* = 10^9$ CFU, and $[Amp_h] = 8 \,\mu M$ ; $[S_p]^* = 6.10^5$ CFU, \\
together with the values of $K_{sc}$ and $K_{sp}$ (from \eqref{Kse} and \eqref{Ksa}) in \eqref{NakaEqSS}, to obtain the following equations:

\begin{System}
\frac{d_{sph}}{r_{sp}} = \frac{4 +C_{ah}}{6} \\ \\
\frac{d_{sph}}{r_{sp}} = \frac{(10^4 - 2)(C_{ah} +8)}{8.10^4}
\end{System}
which reduce to $C_{ah} = 8$ and $\frac{d_{sph}}{r_{sp}} = 2$.\\

Following the same method with the experimental conditions without any LL-37 (i.e. $[Amp_h] = 0$) and using two data points \\
($[Amp_b] = 0.32 \,\mu M$ ; $[S_p]^* = 5.10^8$ CFU) and ($[Amp_b] = 0.64 \; \mu M$ ; $[S_p]^* = 3.10^3$ CFU), \\
we get $C_{ab} = 0.16$ and $\frac{d_{spb}}{r_{sp}} = \frac{5}{4}$.

It is notable that the maximum killing rates of $S_p$ by $Amp_b$ and $Amp_h$ are both proportional to $S_p$ growth rate. Interestingly, such proportional relation has been observed experimentally between the killing rate of \textit{Escherichia coli} by an antibiotic and the bacterial growth rate \cite{tuomanen_rate_1986}.

To be consistent with the ranges of \textit{Sh}-lantibiotics concentrations described in Nakatsuji \textit{et al.} \cite{nakatsuji_antimicrobials_2017}, $[Amp_b]$ should take positive values below 10. Given that $[Amp_b]^* = \frac{k_c [S_c]^*}{d_a}$ at steady-state, and that $K_{sc} = 4.10^8$ CFU is the upper bound for $[S_c]^*$, we obtain the following constraint:

\begin{equation}
    \frac{k_c}{d_a} \leq \frac{1}{4.10^7}
\end{equation}

\subsection{Parameter relations inferred from co-culture data}

The initial model described earlier is representative of the experimental settings of the co-culture conditions described in Kohda \textit{et al.} \cite{kohda_vitro_2021}. At steady-state, the system \eqref{fullSystem} gives:

\begin{equation}\label{SEcoKohda}
[S_c]^* = 0 \quad \textrm{or} \quad [S_c]^* = K_{sc} \left ( 1 - \frac{d_{sc} [S_p]^*}{r_{sc} (C_1 + [S_p]^*)} \right)
\end{equation}
\begin{equation}\label{SAcoKohda}
[S_p]^* = 0 \quad \textrm{or} \quad [S_p]^* = K_{sp} \left ( 1 - \frac{d_{spb} [Amp_b]}{r_{sp} (C_{ab} + [Amp_b])} - \frac{d_{sph} [Amp_h]}{r_{sp} (C_{ah} + [Amp_h])} \right)
\end{equation}
\begin{equation}
{[Amp_b]}^* = \frac{k_c [S_c]^*}{d_a}
\end{equation}

Considering that what is observed experimentally after 48 hours of incubation is at steady-state, one can replace $[S_c]^*$ and $[S_p]^*$ with the experimental data point (\textit{S. epidermidis} $= 10^8$ CFU; \textit{S. aureus} $= 10^9$ CFU) in \eqref{SEcoKohda} and \eqref{SAcoKohda} to get the following parameter relation:

\begin{equation}\label{paramRKohda1}
\frac{d_{sc}}{r_{sc}} = \frac{3}{4.10^9} C_1 + \frac{3}{4}
\end{equation}
\begin{equation}\label{paramRKohda2}
\frac{2}{3} r_{sp} = \frac{d_{sph} [Amp_h]}{C_{ah} + [Amp_h]} + \frac{10^8 d_{spb} k_c}{d_a C_{ab}+10^8 k_c}
\end{equation}

By integrating the values found for $C_{ah}$ and $C_{ab}$, and the relations involving $d_{sph}$ and $d_{spb}$ into \eqref{paramRKohda2}, we end up with:

\begin{equation}\label{relation_da}
    d_a = 10^8 k_c \, \frac{56 + 31 [Amp_h]}{2.56\,(4-[Amp_h])} \quad \textrm{with } [Amp_h] < 4
\end{equation}

\section{Reduced model with 5 parameters} \label{sec:ReducedModel}

Using the previously mentioned experimental data, and assuming they represent steady state conditions of the initial model \eqref{fullSystem}, we have reduced the parametric dimension of the model from 13 to 5. Specifically, out of the original 13 parameters, we could define the values of 4 of them, and derive 4 functional dependencies from the values of the remaining parameters, as summarized in Table \ref{tabReducedModel}).

\begin{table}
\caption{Summary of the parameter relations embedded in the reduced model.}\label{tabReducedModel}
\begin{center}
\begin{tabular}{c|c}
\textbf{Parameter} & \textbf{Value or relation to other parameters}\\
\hline
$K_{sc}$ & $4.10^8$ \\
$K_{sp}$ & $3.10^9$ \\
$C_{ah}$ & 8 \\
$C_{ab}$ & 0.16 \\
$d_{sph}$ & $2 \, r_{sp}$\\
$d_{spb}$ & $\frac{5}{4} \, r_{sp}$\\
$d_{sc}$ & $\displaystyle{r_{sc} \left (\frac{3}{4.10^9} \; C_1 + \frac{3}{4} \right )}$ \\
$d_a$ & $\displaystyle{10^8 k_c \, \frac{56 + 31 [Amp_h]}{2.56\,(4-[Amp_h])}}$ with $[Amp_h] < 4$
\end{tabular}
\end{center}
\end{table}
 
 {In our skin microbiome model \eqref{fullSystem}, the parameters that remain unknown are thus:}
 \begin{itemize}
     \item 
 $r_{sc}$, the growth rate of $S_c$ which can reasonably take values between 0 and 2 $h^{-1}$ 
 following \cite{czock_mechanism-based_2007,campion_pharmacodynamic_2005};
 \item
$r_{sp}$, the growth rate of $S_p$, taking similar values in the interval between 0 and 2 $h^{-1}$;
\item
$C_1$, the concentration of $S_p$ that induces half the maximum killing rate $d_{sc}$ (in $CFU.ASU^{-1}$)
and is thus bounded by the optimum concentration of $S_p$, i.e.~$K_{sp} = 3.10^9 \; CFU.ASU^{-1}$, as calculated in section \ref{subsec:Param_mono} from \cite{kohda_vitro_2021};
\item
$k_c$, the production rate of $[Amp_b]$ chosen to take values between $0$ and $0.1 \; AU.h^{-1}.CFU^{-1}$, and shown to have a limited impact on the steady-state values in section \ref{subsec:Sensitivity};
\item$[Amp_h]$, the concentration in $AU.ASU^{-1}$ of AMPs produced by skin cells between 0 and 4 (equation \eqref{relation_da}).
 \end{itemize}

\subsection{Simulations at the time scale of the experiments}

In order to reproduce what was observed by Kohda et. al \cite{kohda_vitro_2021}, 
that is a dominant pathogenic population after 50 hours 
which can thus be considered as dysbiosis in our skin microbiome model,
it is sufficient to fix a relatively low concentration of Amp produced by the skin cells, i.e. $Amph_h=1.5$,
and some fixed values for the four other parameters chosen in their intervals described above.
Among a continuum of possible solutions, we chose
$r_{sc} = 0.5, \, r_{sp} = 1 , \, C_1 = 5.10^6 , \, k_c = 0.01$.

\begin{figure}
    \centering
    \includegraphics[width=0.8\textwidth]{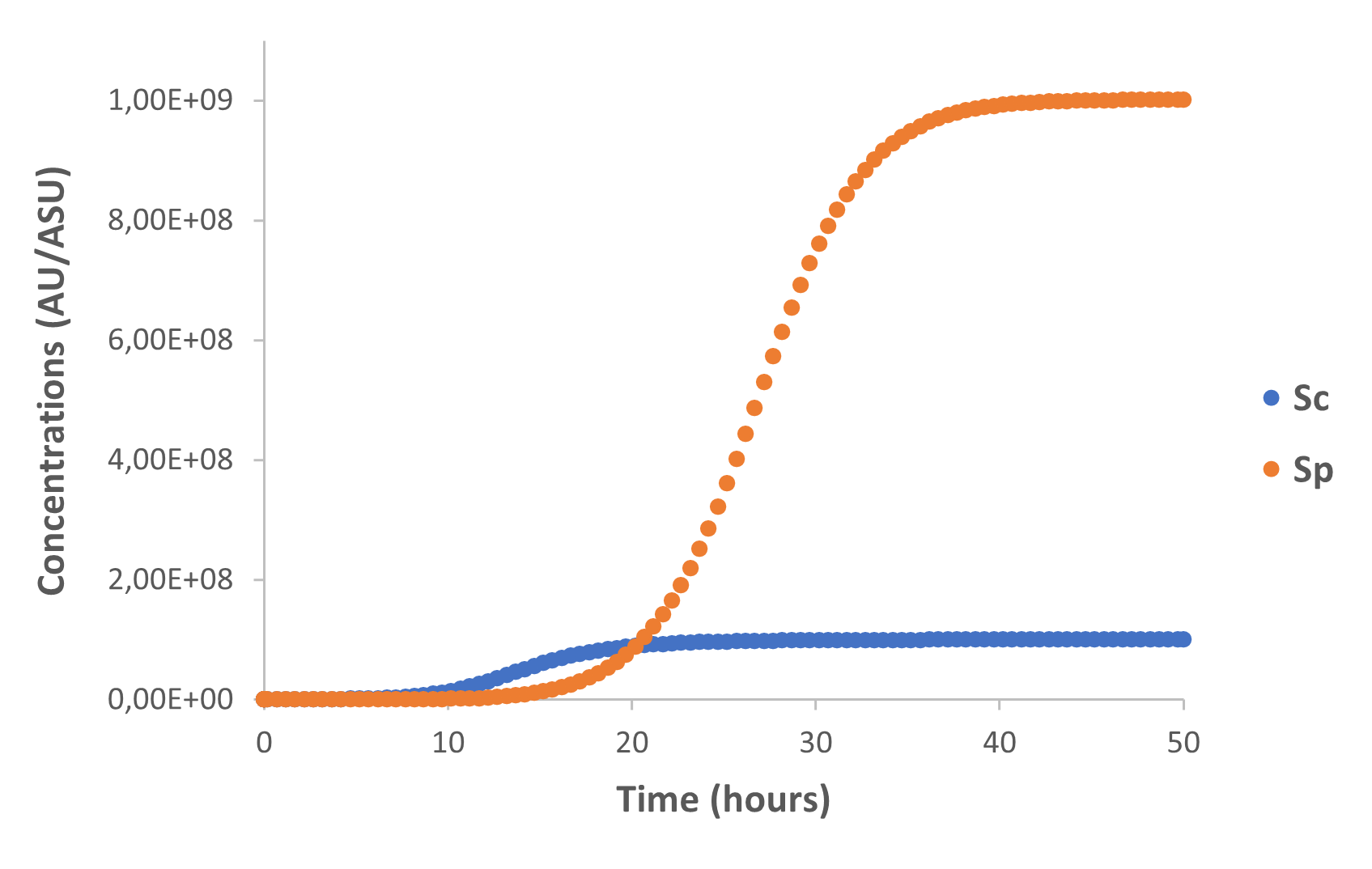}
    \caption{Numerical simulation of the reduced ODE model over 50 hours,
    with initial conditions $[S_c]=10^5,\ [S_p]=10^3,\ [Amp_b]=0$
    and parameter values $[Amp_h]=1.5,\ r_{sc} = 0.5, \, r_{sp} = 1 , \, C_1 = 5.10^6 , \, k_c = 0.01,$ to fit Kohda et al. co-culture data \cite{kohda_vitro_2021} (Table \ref{tabExpDataKohda}).}
    \label{fig:Sim_Kohda}
\end{figure}

The doses of \textit{S. epidermidis} and \textit{S. aureus} applied at the surface of the 3D epidermal equivalent at the beginning of the experiment ($10^5 CFU/mL$ and $10^3 CFU/mL$ respectively) are used as the initial concentrations for $[S_c]$ and $[S_p]$ respectively.
Fig. \ref{fig:Sim_Kohda} shows the result of a numerical simulation\footnote{All computation results presented in this paper have been done using the BIOCHAM software with a notebook runnable online and available at \url{https://lifeware.inria.fr/wiki/Main/Software\#CMSB22b}.} of our model 
with those parameters which are in accordance to the co-culture experiments of Kohda et. al and reproduce a consistent qualitative behavior \cite{kohda_vitro_2021}.

Our model can also be used to reproduce what is considered a balanced microbiome, corresponding to the commensal population being significantly more abundant than  the pathogenic one.
This requires modifying some parameter values to represent a less virulent pathogenic population, closer to the physiological context, given that the experiments from Kohda et. al \cite{kohda_vitro_2021} were performed using a virulent methicillin-resistant \textit{S. aureus} strain.

\begin{figure}
    \centering
    \includegraphics[width=0.8\textwidth]{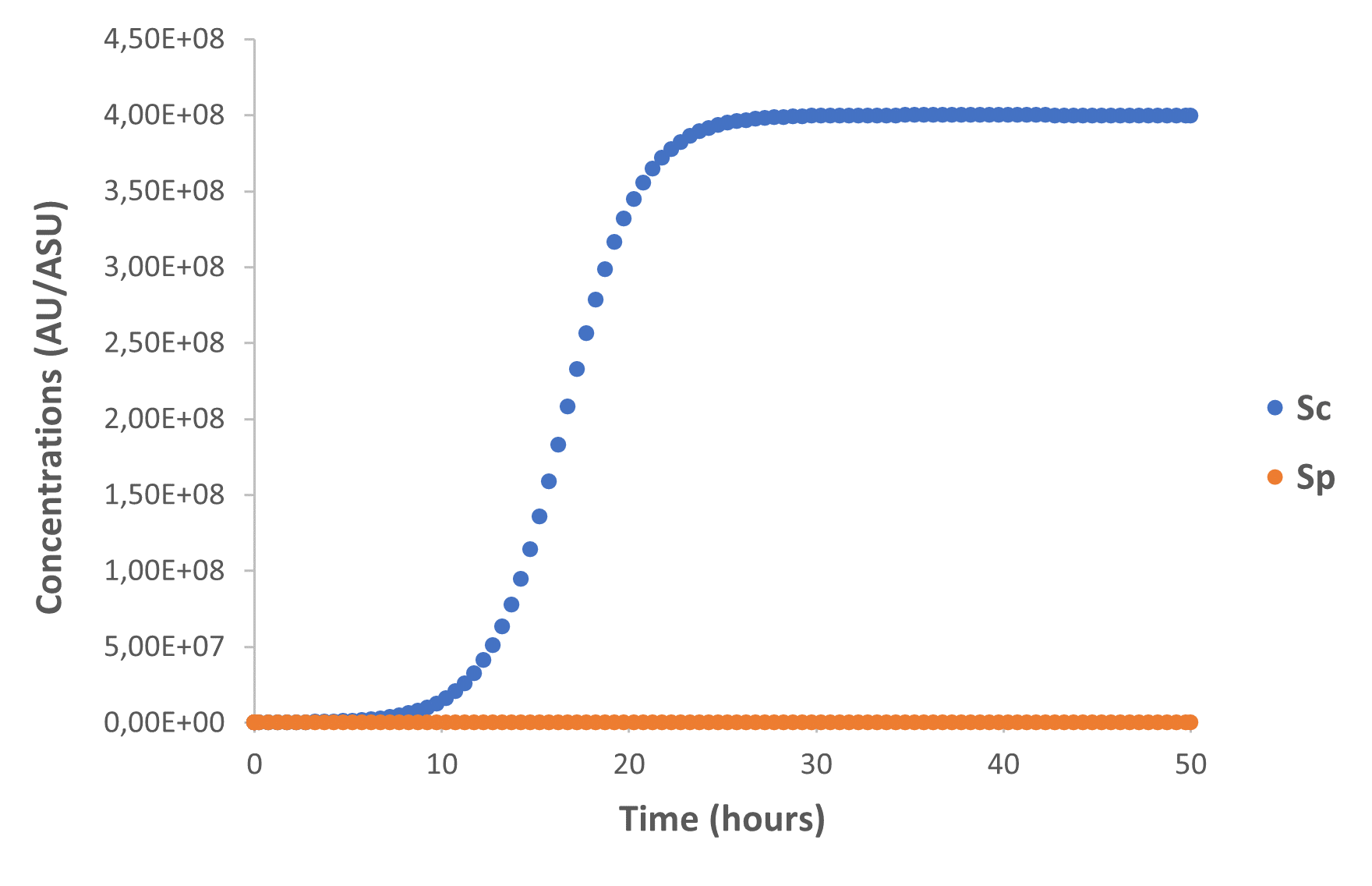}
    \caption{Numerical simulation of the reduced ODE model over 50 hours,
    with initial conditions $[S_c]=10^5,\ [S_p]=10^3,\ [Amp_b]=0$
    and parameter values $r_{sc} = r_{sp} = 0.5 , \, C_1 = 2.10^8 , \, k_c = 0.01 ,\ [Amp_h] = 3$ corresponding to Kohda et al. experiments \cite{kohda_vitro_2021}.}
    \label{fig:Sim_default}
\end{figure}

We chose $r_{sp} = 0.5$, $C_1 = 2.10^8$ and 
a higher production of AMPs by the skin cells, $[Amp_h] = 3$, to compensate for feedback loops or stimuli that might be missing in the 3D epidermal equivalent used.
Fig. \ref{fig:Sim_default} shows a simulation trace obtained  under those conditions
which clearly indicates the dominance of the non-pathogenic population
under those conditions.

\subsection{Parameter sensitivity and robustness analyses} \label{subsec:Sensitivity}

Since the previous simulations rely on some choices of values 
for the unknown parameters, it is important to evaluate the robustness of the predictions of our model
by performing an analysis of sensitivity to the parameter values.
This is possible in Biocham by specifying the property of interest in quantitative temporal logic \cite{RBFS11tcs}.
The interesting property here is  the stabilization at the time scale of the experiments around 48 hours of the bacterial population sizes to the values given by simulation (Fig. \ref{fig:Sim_default}), 
Here we use the temporal logic formula:\\
$F(Time==40 \wedge NSc = x1 \wedge NSp = y1 \wedge F(G(NSc = x2 \wedge NSp = y2)))$\\
and objective values equal to 1 for the free variables $x1, x2, y1, y2$, 
to express that the normalized variables $NSc$ and $NSp$, i.e. current values of $Sc$ and $Sp$ divided by their expected value at steady state, respectively $10^8$ and $10^9$ in the pathogenic case of Kohda et al. experiments, is reached (F, finally) at time around 40 and finally at the end of the time horizon (FG) of 50 hours.
On a given simulation trace, the free variables of the formula have a validity domain (here fixed values) which is used to define a continuous degree of satisfaction of the property as a distance to the objective values, and a robustness degree by sampling parameter values around their nominal values
\cite{RBFS11tcs}.

The sensitivity analysis (Table \ref{tabSensitivity}) reveals that the dominance of the commensal population is highly sensitive to variations of the initial concentration of the pathogen. To a lesser extend, the dominant population is also sensitive to the growth rates ($r_{sc}$ and $r_{sp}$) and the concentration of human AMPs ($[Amp_h]$). On the other hand, $C_1$ and $k_c$ do not seem to affect the relative proportions of the bacterial populations.

\begin{table}
\caption{Sensitivity of the model to variations of the parameters and initial concentrations for the property of reaching the same values at time 40 and time horizon 50 as in Fig.~\ref{fig:Sim_default}.}\label{tabSensitivity}
\begin{center}
\begin{tabular}{c|c|c}
\textbf{Parameter} & \textbf{Coefficient of variation} & \textbf{Robustness degree}\\
\hline
$r_{sc}$ & 0.2 & 0.62\\
$r_{sp}$ & 0.2 & 0.57\\
$C_1$ & 10 & 0.95\\
$k_c$ & 1 & 0.95\\
$[Amp_h]$ & 0.2 & 0.53\\
$[S_p](t=0)$ & 10 & 0.23 \\
$[S_c](t=0)$ & 10 & 0.58\\
$(r_{sc},r_{sp})$ & 0.2 & 0.48\\
$([S_c](t=0),[S_p](t=0))$ & 10 & 0.31\\
\end{tabular}
\end{center}
\end{table}

\subsection{Meta-stability revealed by simulation on a long time scale}

\begin{figure}
    \centering
    \includegraphics[width = 0.8\textwidth]{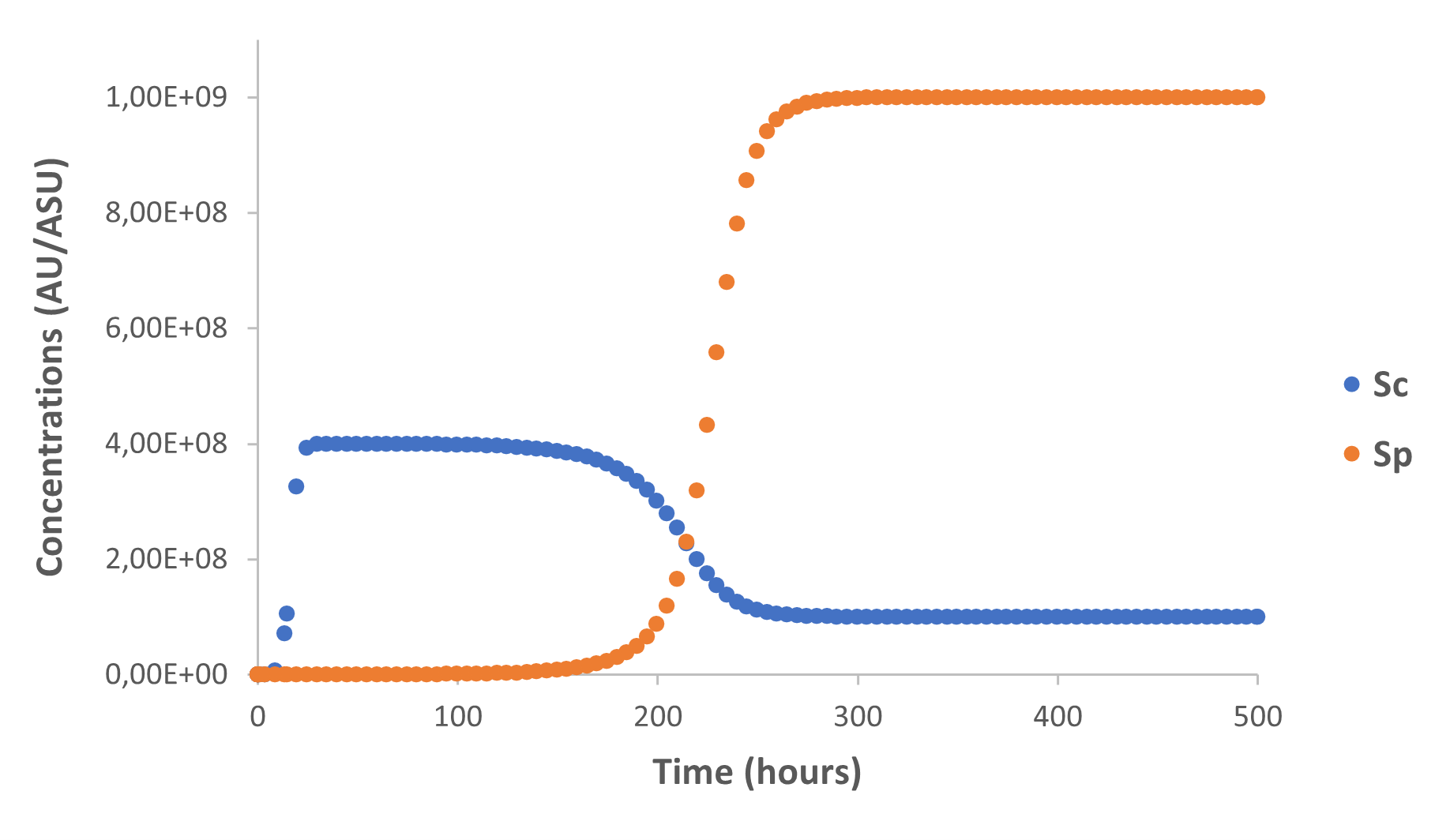}
    \caption{{Numerical simulation of the reduced ODE model on a longer time scale of 500 hours, with the same initial concentrations and parameter values as in Fig.~\ref{fig:Sim_default}, showing an inversion of the dominant bacterial population after 220 hours.}}
    \label{fig:metastab}
\end{figure}

Interestingly, by extending the simulation time horizon to a longer time scale of 500 hours,
one can observe a meta-stability phenomenon, shown in Fig.~\ref{fig:metastab}.
The seemingly stable state observed in Fig.~\ref{fig:Sim_default} at the relevant time scale of 50 hours of the experiments, is thus not a mathematical steady state, 
but a meta-stable state, also called quasi-stable state,
that slowly evolves, with $\frac{d[S_c]}{dt} \neq 0$ and $\frac{d[S_p]}{dt} \neq 0$,
towards a true stable state of the model reached around 300 hours
in which the population density are reversed.

The $S_c$ population almost reaches its optimum capacity $K_{sc}$ after approximately 30 hours and stays relatively stable for around 100 hours more, that is over 4 days, which can reasonably be considered stable on the microbiological time scale. Meanwhile, the $S_p$ population is kept at a low concentration compared to $S_c$, even though it is continuously increasing and eventually leading to its overtake of $S_c$.

By varying the parameters values, it appears that this meta-stability phenomenon emerges above a threshold value of $2.5$ for $[Amp_h]$ \footnote{All computation results presented in this paper have been done using the BIOCHAM software with a notebook runnable online and available at \url{https://lifeware.inria.fr/wiki/Main/Software\#CMSB22b}.}, that is for almost half of its possible values (see section \ref{sec:ReducedModel}). 

That phenomenon of meta-stability, also called quasi-stability, is a classical notion of dynamical systems theory, particularly well-studied in the case of oscillatory systems for which analytical solutions exist, and as models of brain activity \cite{TK08neuron}.
It is worth noting that it has also been considered in the computational systems biology community with respect to model reduction methods based on the identification of different regimes 
corresponding to different preponderant terms of the ODEs, for which simplified dynamics can be defined, and chained within a hybrid automaton \cite{RSNGW15cmsb}.

More generally, this raises the question of the existence and importance of meta-stability in real biological processes, 
as well as the validity of the steady state assumptions made in mathematical modeling methods to fit the models to the observed experimental data.

\section{Conditions favoring the pathogenic population}
Whether the dysbiosis observed in AD is the cause or the result of the disease is unclear \cite{kobayashi_dysbiosis_2015,koh_skin_2021}. Infants developing AD do not necessarily have more \textit{S. aureus} present on their skin prior to the onset of the disease compared to the healthy group \cite{kennedy_skin_2017}. This suggests that atopic skin has some characteristics enabling the dominance of \textit{S. aureus} over the other species of the microbiome. To test this hypothesis, we investigate two changes of the skin properties observed in AD patients (skin surface pH elevation \cite{eberlein-konig_skin_2000} and reduced production of AMPs \cite{ong_endogenous_2002}) and their impact on the dominant species at steady-state. More specifically, we study the behavior of the system following the introduction of a pathogen and whether the pathogen will colonize the media depending on the initial concentrations of the bacterial populations and the particular skin properties mentioned before.

\subsection{Skin surface pH elevation}

According to Proksch \cite{proksch_ph_2018}, the physiological range for skin surface pH is 4.1-5.8. However, in certain skin conditions, like AD, an elevation of this pH has been observed.
Dasgupta \textit{et al.} studied \textit{in vitro} the influence of pH on the growth rates of \textit{S. aureus} and \textit{S. epidermidis}\cite{dasgupta_16502_2020}. Their experimental results show that, when the pH is increased from 5 to 6.5, the growth rate of \textit{S. epidermidis} is multiplied by 1.8, whereas the one of \textit{S. aureus} is multiplied by more than 4 (Table \ref{tabDasgupta}).

\begin{table}
\caption{Experimental data from Dasgupta et. al \cite{dasgupta_16502_2020} showing the influence of pH on growth rates of \textit{S. epidermidis} and \textit{S. aureus}}\label{tabDasgupta}
\begin{center}
\begin{tabular}{c|c|c|}
\multirow{2}{0.6cm}{\textbf{pH}} & \multicolumn{2}{c|}{\textbf{Growth rate ($\Delta$OD/hour)}}\\
\cline{2-3}
& \textit{\textbf{S. aureus}} & \textit{\textbf{S. epidermidis}}\\
\hline
5 & 0.03 & 0.05\\
5.5 & 0.04 & 0.07\\
6 & 0.09 & 0.08\\
6.5 & 0.13 & 0.09\\
7 & 0.14 & 0.10\\

\end{tabular}
\end{center}
\end{table}

Their data can be used to select values for the growth rates $r_{sc}$ and $r_{sp}$ in our model, corresponding to healthy skin with a skin surface pH of 5 and compromised skin with a pH of 6.5.
Because the experiments from Dasgupta \textit{et al.} were performed \textit{in vitro} and the bacterial population sizes measured with optical density (OD) instead of CFU, the growth rates cannot be directly translated into $r_{sc}$ and $r_{sp}$.
We use $r_{sc} = 0.5$ as the reference value for the commensal growth rate at pH 5, following on from previous simulation (Fig. \ref{fig:Sim_default}). Maintaining the ratio between the two population growth rates at pH 5 and the multiplying factors following the pH elevation from Dasgupta \textit{et al.} experimental data, we can define two sets of values for $r_{sc}$ and $r_{sp}$:

\begin{equation*}
\textrm{skin surface pH of 5} \quad \Rightarrow r_{sc} = 0.5, \, r_{sp} = 0.3
\end{equation*}
\begin{equation*}
\textrm{skin surface pH of 6.5} \quad \Rightarrow r_{sc} = 0.9, \, r_{sp} = 1.3
\end{equation*}

Considering the healthy skin scenario with a skin surface pH of 5, the influence of the bacterial populations initial concentrations on the dominant species after 50 hours is evaluate using the temporal logic formula:

$$F(\textrm{Time}==40 \wedge ([S_c] > u1 \, [S_p]) \wedge F(G([S_c] > u2 \, [S_p])))$$

where $u1$ and $u2$ are free variables representing the abundance factors between both populations,
evaluated at Time$= 40$ and at the last time point of the trace respectively (F stands for finally and G for globally at all future time points), i.e. at the time horizon of the experiments of 50 hours.

When given with an objective value, e.g. $u1=10$, the distance between that value
and the validity domain of the formula, i.e. the set of values for $u1$
that satisfy the formula, provides a violation degree which is used to evaluate 
the satisfaction degree of the property.

\begin{figure}
    \centering
    \includegraphics[width = 0.7\textwidth]{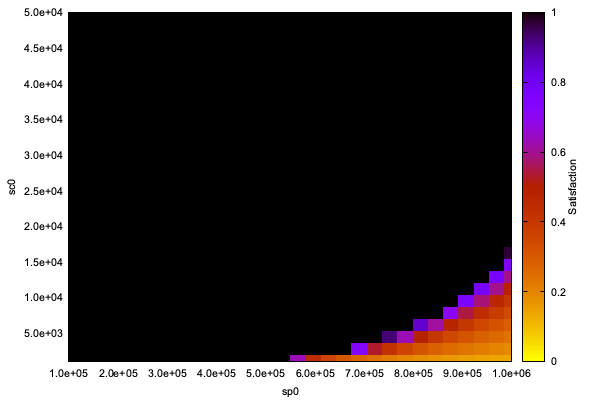}
    \caption{Landscape of satisfaction degree of the temporal formula corresponding to healthy skin with a skin surface pH of 5 ($r_{sc} = 0.5$ and $r_{sp} = 0.3$). The x and y axis represent variations of the initial quantities of $[S_p]$ and $[S_c]$ respectively. The color coding corresponds to the satisfaction degree of the temporal logic formula. Values used for the other parameters: $C_1 = 2.10^8$, $k_c = 0.01$, $[Amp_h] = 3$.}
    \label{fig:landscape_lowpH}
\end{figure}

Here, we evaluate how much the temporal formula $F(\textrm{Time}==40 \wedge ([S_c] > u1 \, [S_p]) \wedge F(G([S_c] > u2 \, [S_p])))$, $u1 \rightarrow 10, \, u2 \rightarrow 10$, is satisfied given variations of the initial concentrations of two populations (Fig. \ref{fig:landscape_lowpH}).
The model predicts that, under the healthy skin condition, the commensal population will always dominate after 50 hours, except when introduced at a relatively low concentration ($<2.10^4$) while the initial concentration of the pathogenic population is high ($>5.10^5$).

\begin{figure}
    \centering
    \includegraphics[width = 0.7\textwidth]{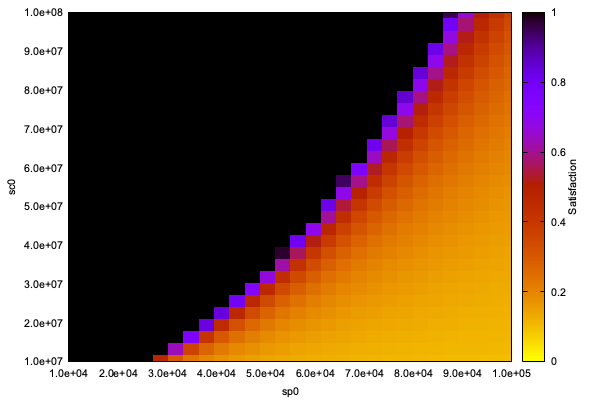}
    \caption{Landscape of satisfaction degree of the temporal formula corresponding to compromised skin with a skin surface pH of 6.5 ($r_{sc} = 0.9$ and $r_{sp} = 1.3$). The x and y axis represent variations of the initial quantities of $[S_p]$ and $[S_c]$ respectively. The color coding corresponds to the satisfaction degree of the temporal logic formula. Values used for the other parameters: $C_1 = 2.10^8$, $k_c = 0.01$, $[Amp_h] = 3$.}
    \label{fig:landscape_highpH}
\end{figure}

The model predicts a higher vulnerability of the skin regarding invading pathogens with an elevated skin surface pH. When evaluating the same temporal formula with growth rates values corresponding to a skin surface pH of 6.5, we observe that even when the initial concentration of commensal is high ($>10^7$), the pathogenic population is able to colonize the skin when introduced at a concentration as low as $3.10^4$ (Fig. \ref{fig:landscape_highpH}).

Such predictions highlight the protective effect of the skin surface acidic pH against the invasion of pathogenic bacteria.

\subsection{Reduced production of skin AMPs}

\begin{figure}
    \centering
    \includegraphics[width = 0.9\textwidth]{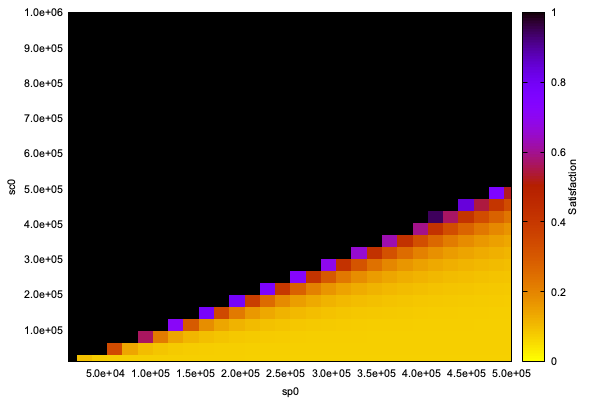}
    \includegraphics[width = 0.9\textwidth]{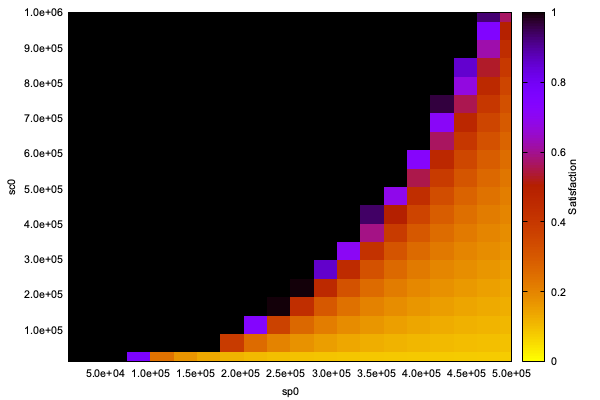}
    \caption{Landscape of satisfaction degree of the healthy condition formula with a low concentration of human AMPs on the upper graph ($[Amp_h] = 0.5$) and a high concentration at the bottom ($[Amp_h] = 3$). The x and y axis represent variations of the initial quantities of $[S_p]$ and $[S_c]$ respectively. The color coding corresponds to the satisfaction degree of the temporal logic formula. Values used for the other parameters: $r_{sc} = r_{sp} = 0.5$, $C_1 = 2.10^8$, $k_c = 0.01$.}
    \label{fig:landscape_Amph}
    
\end{figure}

As mentioned before, human keratinocytes constitutively produce AMPs as a defense against pathogens. In atopic dermatitis, the expression of AMPs is dysregulated, leading to lower concentration levels of AMPs in the epidermis \cite{nakatsuji_antimicrobials_2017}. 
Similarly to the analysis done for skin surface pH, our model can be used to study how the skin microbiome reacts to modulation of the AMPs production by the skin cells.
Two situations are considered: an impaired production of AMPs by the skin cells ($[Amp_h] = 0.5$) and a higher concentration with $[Amp_h] = 3$.
Using the same methodology as in the case of skin surface pH, the temporal logic formula $F(\textrm{Time}==40 \wedge ([S_c] > u1 \, [S_p]) \wedge F(G([S_c] > u2 \, [S_p])))$, $u1 \rightarrow 10, \, u2 \rightarrow 10$, is evaluated for variations of the initial concentrations of both populations for $[Amp_h] = 0.5$ and $[Amp_h] = 3$ (Fig. \ref{fig:landscape_Amph}).\\

The model predicts a slightly protective effect of $Amp_h$ regarding the colonization of the skin by a pathogenic population, for low initial concentrations. However when both populations are introduced in high concentrations, the increase of $[Amp_h]$ appears to have the opposite effect of facilitating the colonization by the pathogenic population.

This mitigated effect might be due to the presence of $[Amp_h]$ in the constraint related to the degradation rate of $[Amp_b]$ (equation \eqref{relation_da}) and deserves further investigation.

\section{Conclusion}

The objective of this research is the identification of conditions which might favor or inhibit the emergence of pathogenic populations in the skin microbiome. Such analyses can lead to insights about potential treatment strategies aiming at restoring a dysbiotic condition.

We have developed a simple ODE model of skin microbiome with 3 variables and 13 parameters
which could be reduced to 5 parameters by using published data from the literature
and steady state reasoning on the observations made in the biological experiments.
Our bacterial population model is generic in the sense that we did not 
take into account the peculiarities of some specific bacterial populations,
but on some general formulas of adversary population dynamics and influence factors.
We showed through sensitivity analyses that our model predictions are particularly robust
with respect to parameter variations.

Perhaps surprisingly, we also showed that this simple model exhibits over a large range of biologically relevant parameter values, a meta-stability phenomenon, revealed by allowing the simulation  to continue for times one order of magnitude longer than the reported experimental times.
This observation questions the existence and importance of meta-stability phenomena 
in real biological processes, whereas a natural assumption made in mathematical modeling, 
and model fitting to data, is that the experimental data are observed in states corresponding to real stable states of the mathematical model.

\subsubsection*{Acknowledgments.} We are grateful to Mathieu Hemery, Aurélien Naldi and Sylvain Soliman for interesting discussions on this work.

\bibliographystyle{splncs04}
\bibliography{CMSB_2022}

\end{document}